\documentclass{article}

\PassOptionsToPackage{numbers, compress}{natbib}
\usepackage[final, main]{neurips_2026}

\usepackage[utf8]{inputenc} %
\usepackage[T1]{fontenc}    %
\usepackage{url}            %
\usepackage{booktabs}       %
\usepackage{amsfonts}       %
\usepackage{amsmath}
\usepackage{amssymb} %
\usepackage{amsthm}
\usepackage{stmaryrd}
\usepackage{nicefrac}       %
\usepackage{microtype}      %
\usepackage[dvipsnames]{xcolor}         %
\usepackage{algorithm}
\usepackage{algpseudocode}
\usepackage{caption}
\usepackage{listings}

\usepackage[frozencache,cachedir=.]{minted}
\usepackage{mathtools}
\usepackage{multirow}
\usepackage{multicol}
\usepackage{tcolorbox}
\usepackage{dsfont}
\usepackage{xspace}
\usepackage{blkarray}
\usepackage{pgfplots}
\usepackage{hyperref}
\usepackage{tabularx}
\usepackage{booktabs}
\pgfplotsset{compat=1.18}
\usepackage{pgfplotstable}
\usepackage[inline]{enumitem}
\usepackage{wrapfig}
\usepackage{needspace}

\usepackage{tikz}
\usetikzlibrary{automata, positioning, arrows.meta, calc, shapes.geometric}

\usepackage[capitalize,noabbrev]{cleveref}

\theoremstyle{plain}
\newtheorem{theorem}{Theorem}[section]

\theoremstyle{definition}

\theoremstyle{remark}

\makeatletter
\newtheorem*{rep@theorem}{\rep@title}
\newcommand{\newreptheorem}[2]{%
\newenvironment{rep#1}[1]{%
 \def\rep@title{#2 \ref{##1}}%
 \begin{rep@theorem}}%
 {\end{rep@theorem}}}
\makeatother

\newcommand{\tname}[1]{\textsc{#1}\xspace}
\newcommand{\tool}{\tname{Elba}}

\setmintedinline{formatcom=\let\PYG\plaincodetoken}
\newmintinline[code]{haskell}{}
\newcommand{\agent}{\code{agent}}

\usetikzlibrary{arrows.meta,positioning,fit,calc,shapes.geometric}

\title{Language-Based Agent Control}

\author{%
  Timothy Zhou, Loris D'Antoni, Nadia Polikarpova \\
  Department of Computer Science and Engineering \\
  University of California-San Diego \\
  La Jolla, CA 92037 \\
}

\begin{document}
\raggedbottom

\maketitle

\begin{abstract}
This paper introduces \emph{language-based agent control} (LBAC),
a new programming model for agentic applications that brings techniques from programming languages
and language-based security to the problem of agent control.
In conventional programming, combinations of static typing and runtime enforcement have long been used to guarantee that well-typed programs satisfy user-specified \emph{policies}, including policies for access control, information flow, data provenance, and more.
The key idea behind LBAC is to extend these guarantees to agentic applications by \textit{requiring agents to generate programs that are themselves {well-typed}} in the context of the surrounding scaffolding code.
Unsafe programs are rejected by the typechecker before execution, allowing policies to apply uniformly across the entire application, including both agent-generated behavior and developer-written scaffolding.
At the same time, LBAC preserves substantial expressiveness:
agents may perform arbitrary side-effect-free computation and recursively invoke subagents, which retain full tool access subject to the same---or potentially more restrictive---policies.
We demonstrate LBAC with three case studies: I/O sandboxing via filesystem capabilities, data provenance, and information-flow control.
\end{abstract}

\section{Introduction}
\label{sec:introduction}

Mainstream agent frameworks~\citep{openai2025agentssdk,anthropic2025agentdocs,langgraph2024,wu2023autogenenablingnextgenllm,crewai2024} force programmers to trade off \textbf{expressiveness} against \textbf{control}.
\textit{Tool}-based systems offer strong \textit{control} but limited \textit{expressiveness}
by restricting the agent to a fixed set of actions, making safety and policy enforcement easier, but limiting composition and arbitrary computation.
In contrast, sandboxed interpreters---such as Python or shell environments---maximize \textit{expressiveness} by allowing agents to write programs, but only \textit{control} coarse-grained system policies (e.g., file-system or network access).
Such interpreters struggle to enforce application-level policies
such as data provenance (``all entries you add to my bibliography must come from a trusted database''),
information flow (``data from customer A should never be sent to customer B''),
or policies scoped to sub-tasks rather than the entire agent (``spawn one sub-agent per customer; each sub-agent only has access to the data owned by that customer'').

We introduce a new programming model for agentic applications,
\emph{language-based agent control} (LBAC), that reconciles \textit{strong control} with \textit{high expressiveness}.
The key observation is that programming languages research has long {used \emph{types} to rule out unsafe behavior in programs}~\citep{DBLP:conf/icfp/VazouSJVJ14,DBLP:journals/pacmpl/BernardyBNJS18,DBLP:journals/jsac/SabelfeldM03}.
{The central claim of LBAC is:}
\begin{center}
{\emph{Type systems can enforce the safety of code with agents.}}
\end{center}

The key idea of LBAC is to \emph{embed agents into a pure, type-safe programming language}%
\footnote{A pure language has no side effects: that is, all the effects a program might perform are reflected in its type.
We illustrate this concept in \autoref{sec:example}.
{Common examples of pure languages include Haskell and Lean.}}, such that
\begin{enumerate*}[label=(\arabic*)]
\item the agent communicates with the world by writing code in the language;
\item the scaffolding code that comprises the agentic application is \emph{also} written in the same language;
\item policies are specified using type signatures.
\end{enumerate*}
{Unlike approaches that require agents to generate all code upfront~\citep{debenedetti2025defeatingpromptinjectionsdesign,DBLP:journals/cacm/Meijer26}, LBAC allows agents to dynamically issue tool calls in response to data.}
{Other approaches~\citep{costa2025securingaiagentsinformationflow,kolluri2026optimizingagentplanningsecurity,stanley2026aiagentexecutionenvironment} rely on custom harnesses built around a specific policy enforcement mechanism---by contrast, LBAC enables existing policy enforcement mechanisms designed for ordinary code to be natively reused for agents.}
Thus, our approach naturally supports a wide range of policies,
including fine-grained capabilities (e.g., file-system or network access),
data provenance and information-flow constraints,
and per-task policies that restrict individual subcomputations.
{LBAC is the first approach to make the agent implementation portable across existing language-based policy enforcement mechanisms.}
{Moreover, the guarantees provided by these mechanisms extend to the entire agentic application.}

\paragraph{Contributions.}
We make the following contributions:

\emph{(1) A new programming model for agentic applications}, \emph{language-based agent control} (LBAC),
which reconciles strong control with high expressiveness by embedding agents in a pure, typed programming language
and expressing policies as types over programs (\autoref{sec:programming-model}).

\emph{(2) \tool, a practical realization of LBAC}, implemented in Haskell.

\emph{(3) Case studies and evaluation} demonstrating the breadth and effectiveness of LBAC.
We instantiate LBAC for three policies: data provenance, file-system capabilities,  and information flow control (IFC).
Our IFC case study notably reuses the mature LIO library~\cite{DBLP:conf/nordsec/StefanRMM11,DBLP:journals/jfp/StefanMMR17}, inheriting its guarantees and illustrating LBAC’s ability to build on prior programming languages research.
{{}Across a range of models and attack settings, LBAC achieves higher task utility on AgentDojo~\citep{debenedetti2024agentdojodynamicenvironmentevaluate} than existing systems such as CaMeL, while providing stronger guarantees across the full agentic loop (\autoref{sec:evaluation}).}

\section{Motivating Example}\label{sec:example}

We start by motivating LBAC through a simple example
that positions it relative to the mainstream approaches---sandboxed interpreters and restricted tools.

{
{}
As a running example, suppose a programmer is implementing a literature search agent that produces BibTeX bibliographies based on a user request. The programmer wants to enforce the following constraint: \emph{every BibTeX file the agent outputs should contain only data sourced from DBLP}. To accomplish this task effectively, the agent needs to be able to (1) inspect bibliography data and (2) apply appropriate computations over the data (e.g., sorting by year {or filtering by a venue}).
}

\begin{figure}[t]
\centering
\begingroup
\setmintedinline{formatcom={}}%
\providecommand{\warn}{\textcolor{red}{$\blacktriangle$}}
\providecommand{\blocked}{%
  \tikz[baseline=-0.5ex]\node[
    regular polygon, regular polygon sides=3,
    draw=black, line width=0.4pt, fill=yellow!80,
    inner sep=0pt, minimum size=1ex]{\textbf{\scriptsize !}};%
}
\providecommand{\ok}{\textcolor{ForestGreen}{\checkmark}}
\providecommand{\bad}[1]{\textcolor{red}{#1}}
\providecommand{\strike}[1]{%
  \tikz[baseline]{\node[inner sep=0pt, outer sep=0pt] (X) {#1};
                  \draw[red, thick, -] (X.west) -- (X.east);}}

\resizebox{\linewidth}{!}{%
\begin{tikzpicture}[
    font=\small,
    >=Stealth,
    node distance=1.4mm and 4mm,
    prompt/.style={draw, rounded corners=6pt, fill=yellow!15,
                   align=left, inner sep=4pt, text width=36mm},
    libbox/.style={draw, rounded corners=3pt, align=left, inner sep=4pt, font=\scriptsize},
    label/.style={font=\small\bfseries},
    agent/.style={draw, rounded corners=2pt, fill=blue!8,
                  align=left, inner sep=3pt, text width=46mm,
                  font=\scriptsize\ttfamily},
    env/.style={draw, rounded corners=2pt, fill=gray!12,
                align=left, inner sep=3pt, text width=46mm,
                font=\scriptsize\ttfamily},
    plain/.style={draw=none, fill=none},
    flow/.style={->, thick, gray!70},
    typeflow/.style={->, thick, gray!70, dashed},
]

\node[libbox, fill=white, align=right] (libnames) at (4.7, 0.4)
{%
  \textbf{Library}\\[1pt]
  \code|dblpSearch|\\
  \code|dblpFetchBib|\\
  \code|appendToBibFile|
};

\node[libbox, anchor=west, fill=blue!4,
      draw=blue!50, dashed]
      (libtypes) at ($(libnames.east) + (0mm,0)$)
{%
  \textbf{\textcolor{blue!60!black}{+ types (LBAC only):}}\\[1pt]
  \code|:: String -> BibIO [DOI]|\\
  \code|:: DOI -> BibIO (Trusted Bib)|\\
  \code|:: FilePath -> Trusted Bib -> BibIO ()|
};

\node[prompt, anchor=east] (prompt) at ($(libnames.west) + (-4mm, 0)$)
  {``Find the earliest paper on differential privacy and add it to \texttt{refs.bib}.''};

\begin{scope}[shift={($(prompt.west) + (-0.45, -0.25)$)}, line width=0.7pt]
  \draw (0,0.55) circle (0.10);                      %
  \draw (0,0.45) -- (0,0.10);                        %
  \draw (-0.13,0.30) -- (0.13,0.30);                 %
  \draw (0,0.10) -- (-0.10,-0.10);                   %
  \draw (0,0.10) -- (0.10,-0.10);                    %
\end{scope}

\node[label] (h1) at (-0.5,-1.35) {\textsf{Code interpreter}};
\node[label] (h2) at ( 4.7,-1.35) {\textsf{Tools}};
\node[label] (h3) at ( 9.9,-1.35) {\textsf{LBAC (\tool)}};

\node[agent, below=2mm of h2.south, anchor=north] (t1) {%
dblpSearch("differential privacy")};
\node[env, below=of t1] (t2) {%
["10.1007", ...]};
\node[agent, below=of t2] (t3) {%
dblpFetchBib("10.1007")};
\node[env, below=of t3] (t4) {%
"@inproceedings\{Dwork2006,\\
\hspace*{2mm}title=\{Calibrating noise...\},\\
\hspace*{2mm}year=\{2006\}, ...\}"};
\node[plain, below=of t4] (t5) {...};
\node[agent, below=of t5] (t6) {%
fetchAndAppend("refs.bib","10.1007")};
\node[env, below=of t6] (t7) {%
\ok\ trusted entry written.\\
\warn\ cannot compute earliest paper\\
\warn\ awkward tool design};

\node[agent, below=2mm of h1.south, anchor=north] (c1) {%
\code|do -- find relevant DOIs|\\
\code|   hits <- dblpSearch "diff. priv."|\\
\code|   -- get bib entry for each DOI|\\
\code|   bibs <- mapM dblpFetchBib hits|\\
\code|   -- find earliest|\\
\code|   return (minimumBy getDate hits)|};

\node[env, below=of c1] (c2) {%
"@inproceedings\{Dwork2006,\\
\hspace*{2mm}title=\{Calibrating noise...\},\\
\hspace*{2mm}year=\{2006\}, ...\}"};

\node[agent, below=of c2] (c3) {%
\code|-- use writeFile from std lib|
\code|writeFile("refs.bib",|\\
\hspace*{2mm}\bad{"@inproceedings\{Dwork06,}\\
\hspace*{4mm}\bad{title=\{Algorithmic...\},}\\
\hspace*{4mm}\bad{year=\{2006\}\}"})};

\node[env, below=of c3] (c3) {%
\warn\ used \code|writeFile| from the\\
\ \ standard library to write a\\
\ \ fabricated bib entry to the file.};

\node[agent, below=2mm of h3.south, anchor=north] (l1) {%
\code|do -- find relevant DOIs|\\
\code|   hits <- dblpSearch "diff. priv."|\\
\code|   -- get bib entry for each DOI|\\
\code|   bibs <- mapM dblpFetchBib hits|\\
\code|   -- write earliest entry to file|\\
\code|   appendToBibFile "refs.bib"|\\
\code|      (minimumBy getDate hits)|};

\node[env, below=of l1] (l2) {%
\ok\ trusted entry written.\\
\ok\ \code|writeFile| would not typecheck\\
\ \ \ (does not return \code|BibIO|)};

\draw[flow] (libnames.south)      to[bend right=10] (h1.north);
\draw[flow] (libnames.south) to[bend right=5]
    node[midway, left=1mm, font=\scriptsize, align=right]
        {\strike{\code|appendToBibFile|}\\\code|fetchAndAppend|}
    (h2.north);
\draw[flow] (libnames.south)      to[bend left=5]   (h3.north);
\draw[typeflow] (libtypes.south)  to[bend left=10]  (h3.north);

\end{tikzpicture}
}
\endgroup
\caption{Implementing a literature search agent using a code interpreter, specialized tools, and \tool (our approach).
\textbf{Sandboxed interpreters} (left panel) enable expressive program composition,
but allow the agent to bypass the library and fabricate entries using the raw file-write function from the standard library.
Restricted \textbf{tools} (middle) enforce strong control by restricting access to trusted operations,
but limit expressiveness; our example requires modifying the set of tools so that the only tool we provide for writing to file also fetches the bib entry directly from the database.
\textbf{\tool} (right) reconciles this tradeoff:
the agent can write programs and compose operations freely,
while the type system enforces that all written entries originate from trusted sources.
}
\label{fig:teaser}
\end{figure}
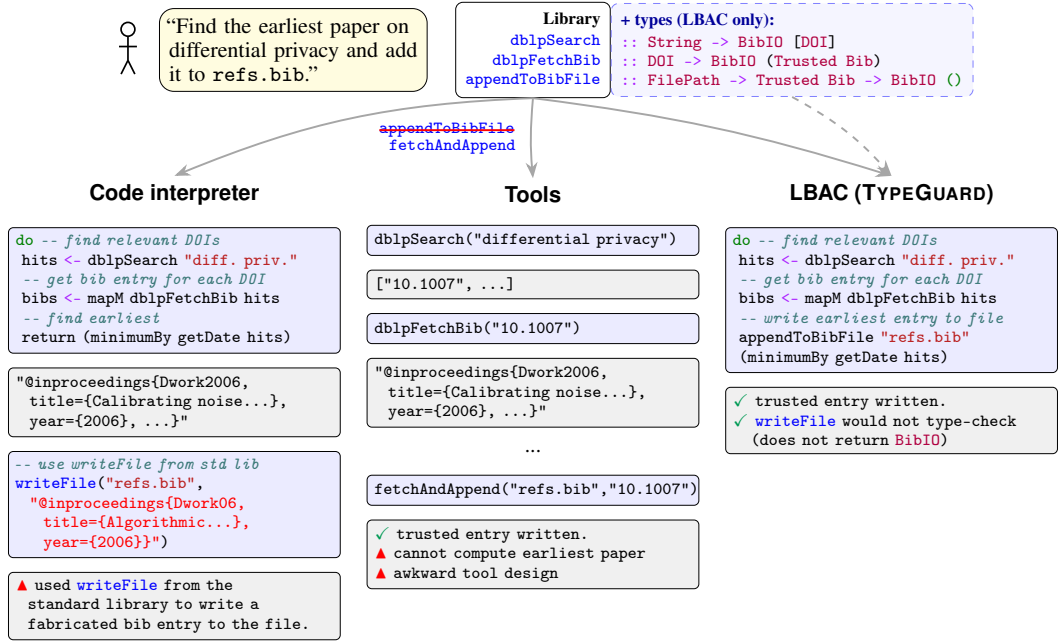

To support this task, the programmer implements a small BibTeX-manipulation library:
\code|dblpSearch| returns the DOIs of papers matching a request,
\code|dblpFetchBib| retrieves the full BibTeX entry for a given DOI,
and \code|appendToBibFile| appends an entry to a file
(\autoref{fig:teaser}, top).
For now, ignore the type signatures next to each function;
they will only come into play once we present LBAC.

\paragraph{Code interpreter: expressive, but unconstrained.}
A natural and expressive way to expose this library is through a sandboxed code interpreter,
in which the agent generates and executes arbitrary code that calls these functions.
The leftmost panel of \autoref{fig:teaser} shows what such a session might look like:
the agent writes a small program that searches DOIs, fetches their entries, and selects the earliest one.
Throughout this paper, we use Haskell syntax for code snippets for consistency,
but for this code-interpreter baseline the agent-generated code should be read with Python-like semantics: dynamically typed, with unrestricted I/O.
Consequently, nothing prevents the agent from forgoing our safe \code|appendToBibFile| function
and instead calling \code|writeFile| from the standard library directly,
with a BibTeX string it constructed itself, fabricating an entry that never came from DBLP.

\paragraph{Restricted tools: safe, but inflexible.}
We could instead enforce stricter control by exposing \emph{only} our BibTeX library as separate tools,
with no access to a code interpreter.
To prevent fabrication, we replace \code|appendToBibFile| with a coarser-grained tool,
\code|fetchAndAppend|, that fetches a DOI and writes its entry to the file in a single {combined} step;
\code|writeFile| is no longer available.
The middle panel of \autoref{fig:teaser} shows the result.
This design indeed ensures that all written entries come from DBLP,
but at the cost of expressiveness:
the agent can no longer use code to compute over results,
and must instead rely on the model to inspect entries one by one in order to find the earliest one.

\newpage

\paragraph{LBAC: expressiveness \emph{and} control.}
Language-Based Agent Control (LBAC) {is a new approach that uses properties of programming languages to enforce policies on agent behavior.}
{A programmer using \tool implements the literature search agent by defining an \emph{embedded domain-specific language} (EDSL) in which the agent expresses its operations.}
{An EDSL is a library which uses language features to enforce policies on programs using the library.
The EDSL acts as a controlled interface between the agent and the tools: for our literature agent, the BibTeX library's type signatures are designed so that the type system enforces the policy that all written entries originate from DBLP.}
{The programmer uses two features of Haskell to enforce the policy:}

\begin{itemize}
\item {\textit{Abstract data types.}}
{An abstract data type is a type whose representation is hidden from code outside its defining module.}
Abstract data types let us define an opaque wrapper type \code|Trusted| whose constructor is hidden inside the library,
and hence \code|Trusted| objects cannot be constructed {or modified directly} by the agent.
{The} only way for the agent to obtain a value of {type \code|Trusted Bib|} is by calling \code|dblpFetchBib|.
Because \code|appendToBibFile| expects a \code|Trusted Bib|,
{the function accepts only records fetched from DBLP.}

\item {\textit{Purity.}}
{Haskell uses types to restrict which effects a computation can perform.
An \emph{effect} is an interaction with external state, such as reading or writing a file.}
Standard-library functions such as \code|writeFile| carry the type \code|IO|, which permits{, for example,} arbitrary {input/output operations such as file access and network communication}.
Our programmer instead defines a custom effect type \code|BibIO|,
{so the \code|IO| computation produced by \code|writeFile| cannot be used as a \code|BibIO|.}

The final ingredient is to require the agent's program to live within this EDSL,
which the programmer does by setting the target type of the agent's response to \code|BibIO ()|.
As a result, any program that combines our three library functions with arbitrary effect-free code{,} as in \autoref{fig:teaser} {(right),} typechecks,
while any attempt to call \code|writeFile| anywhere in the program{,} as in \autoref{fig:teaser} {(left),}
fails to typecheck and is rejected by \tool.
\end{itemize}

\section{LBAC Approach}
\label{sec:programming-model}
This section describes the LBAC programming model.
While the examples in this section use our Haskell implementation, \tool,
the results of this section apply to any pure typed host language.

\autoref{fig:lbac-overview} gives an overview of the LBAC approach,
using the running example from \autoref{sec:example}.
In this programming model, the developer writes their application (the so-called ``scaffolding'') in a pure, typed language, such as Haskell,
using an LBAC library, such as \tool.
The library exposes a single function,
\text{\code{agent :: Defs -> Prompt -> t}},
that takes as input
\begin{enumerate*}[label=(\arabic*)]
    \item \code{Defs}: a set of imports (types and functions) the agent has access to;
    \item \code{Prompt}: a string representing the user prompt;
\end{enumerate*}
and returns an output with an {\emph{arbitrary}} output type \code{t},
which is determined by the context in which \code{agent} is called.
In \autoref{fig:lbac-overview}, the programmer calls \code{agent} with
\begin{enumerate*}[label=(\arabic*)]
    \item the types and functions from our BibTex library (\code{DOI}, \code{Bib}, \code{BibIO}, \code{dblpSearch}, \code{dblpFetchBib}, etc.);
    \item the prompt that includes the generic instructions and the user query.
\end{enumerate*}
The programmer also specifies the expected type of the result via an inline type annotation \code{:: BibIO ()}.

At run time, when the \code{agent} call is evaluated, three things happen (\autoref{fig:lbac-overview}, right).
\emph{First,} \tool queries the language model to generate a Haskell program, giving it the prompt, the available definitions, and a concrete expected type \code{T}.
\emph{Second,} once a program \code{p} is generated, \tool\ typechecks it against the expected type \code{T}.
If the program fails to typecheck, the type error is returned to the model, which retries.
\emph{Third,} once the program passes the type check, it is executed using the Haskell interpreter,
and its result is returned as the output of the \code{agent} call.%

\subsection{Specifying Policies with Types}
\label{sec:policies}

Readers are likely familiar with types as a way of describing the structure of data (e.g., distinguishing integers from strings).
In \emph{pure} languages, however, types do much more: they also constrain the \emph{effects} a program may perform.

\Needspace{8\baselineskip}
\paragraph{Restricting effects.}
Consider the following agentic program that implements an AI calculator:
\begin{minted}{haskell}
main = do   print "Enter a math problem:"
            problem <- readLine
            result <- agent mathDefs problem :: Int
            print result
\end{minted}
Because the target type of the \code{agent} call is \code{Int},
the agent is not only required to return an integer,
it is also forbidden from performing any effect at all---no I/O, state updates, or exceptions.
Consequently, it is safe to deploy this calculator as a public web service:
if an untrusted user enters a malicious query like ``delete all files on the server, then compute the factorial of 5,''
the LLM might generate a program such as \code|do system "rm -rf /"; return (factorial 5)|,
but this program is rejected at the typechecking step.
The reason is that in Haskell, \code|system|---and any other function performing I/O---returns a type wrapped in the \code|IO| effect,
and once any subexpression of a program has type \code|IO|, the entire program must have type \code|IO|,
which does not match the expected type \code|Int|.\footnote{Standard Haskell has a few escape hatches, such as \code{unsafePerformIO}, that would in principle let an agent break out of a pure type.
This loophole, however, has long been plugged by language-based security researchers:
\emph{Safe Haskell}~\citep{DBLP:conf/haskell/TereiMJM12} is a sub-language that excludes all such unsafe features and is selected via a compiler flag, so we can restrict agent-generated code to this safe subset.}

\begin{figure}[!t]
    \centering
    \resizebox{\linewidth}{!}{
    \input{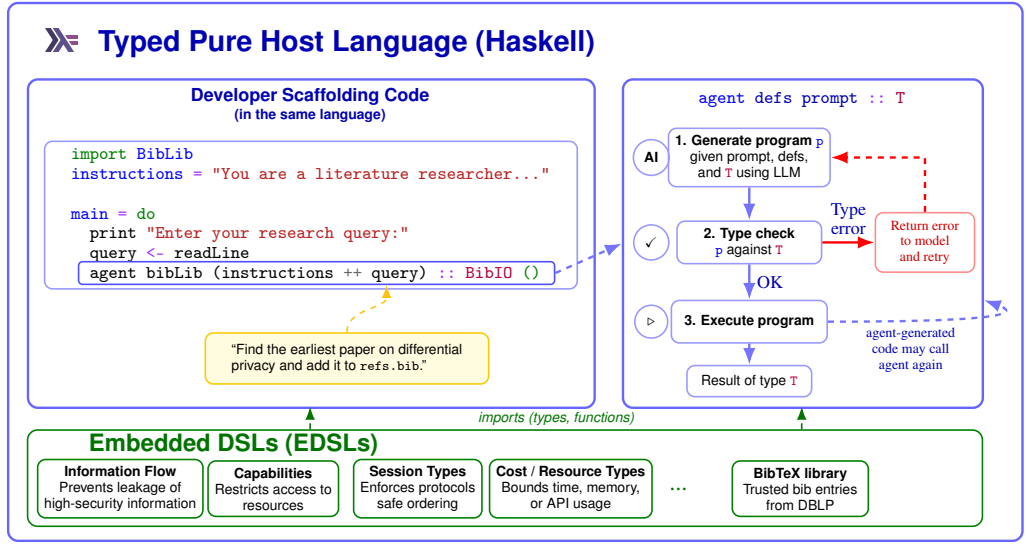}
    }
    \caption{Overview of the Language-Based Agent Control (LBAC) model.
The entire agentic system is expressed within a single typed host language (in the case of \tool, Haskell):
developer-written scaffolding, agent-generated programs, and library code all live in the same environment.
An agent call generates code, which must typecheck before execution.
As a result, the type system acts as the outer layer of control, ensuring that any property it enforces
(e.g., capabilities, information-flow control, or resource bounds) holds for the entire system.
Because \tool is built in Haskell, it can directly leverage the rich ecosystem of existing type systems.}
    \label{fig:lbac-overview}
\end{figure}

\paragraph{Customizing effects with EDSLs.}
Of course, an agent that can compute but cannot perform \emph{any} effect is rather boring;
our bibliography agent, for instance, needs to fetch papers from DBLP and append entries to a file.
The standard programming-languages tool for this situation is an \emph{embedded domain-specific language} (EDSL):
a library that defines custom effect types together with the operations that may produce them,
thereby exposing only those side effects sanctioned by the policy~\citep{DBLP:conf/esop/PlotkinP09,DBLP:conf/popl/Leijen17}.
The bottom of \autoref{fig:lbac-overview} illustrates this pattern:
the BibTeX EDSL from \autoref{sec:example} introduces a custom effect \code{BibIO} together with three operations,
and combines it with the abstract data type \code{Trusted Bib} to encode the provenance policy.
{The} effect type controls \emph{what} the agent can do (only the three sanctioned operations, plus anything effect-free),
while the abstract data type controls \emph{which values} can flow into those operations
(only entries that originated from DBLP).

\paragraph{Reusing existing EDSLs.}
A programmer using \tool can either define their own EDSL---as we did with \code{BibIO}---%
or reuse existing libraries that encode policies as types{.}
{In general, language-based techniques rely on a combination of type system features (Haskell supports many through its extension system~\citep{ghclanguageextensions}) and runtime checks.
Examples from the literature include sandboxing untrusted code~\citep{DBLP:conf/haskell/TereiMJM12}, capability-based access control~\citep{miller2006robustcomposition}, enforcing that operations follow a state-machine protocol~\citep{pucella2008session}, avoiding data races~\citep{harris2005transactions}, and enforcing predicates on values~\citep{DBLP:conf/icfp/VazouSJVJ14}.}
\autoref{sec:evaluation} substantiates this claim
by reusing the LIO library for information-flow control~\citep{DBLP:journals/jfp/StefanMMR17}
completely off the shelf.

\subsection{Programming with Agents}
\label{sec:programming}

Because in LBAC, agents are ordinary functions,
we can use the full power of functional programming---%
higher-order functions, recursion, and type inference---%
to compose agents into data-processing pipelines and multi-step workflows,
\emph{without stepping outside the type system}.

\paragraph{Composing agents.}
Higher-order functions make it easy to plug agents into standard combinators.
For example, the following program summarizes a long document chunk by chunk {using Haskell's \code{foldr} (reduce) function}:
\begin{minted}{haskell}
let summarize chunk summary = agent defs ("Extend this summary: " ++ summary
                                ++ "\nwith a summary of: " ++ chunk)
in foldr summarize "" (chunks document)
\end{minted}
Note that we did not need to annotate the expected type of the \code{agent} call:
because the seed argument of \code{foldr} is the empty string,
Haskell infers that \code{summarize} must return a \code{String},
and \tool\ communicates this inferred type to the model.

This automatic type propagation is more than a convenience:
it lets the developer write short prompts that would otherwise be ambiguous.
Consider:
\begin{minted}{haskell}
let f s = agent defs ("parse this: " ++ s) in
sortBy compareByYear (map f rawDates)
\end{minted}
``Parse'' as what---a date, a timestamp, a duration?
Without type information, the LLM would have to guess from the surrounding prose.
But Haskell's Hindley--Milner type inference deduces \code{f :: String -> Date} from \code{compareByYear},
and \tool\ communicates this expected type to the model,
so the agent knows it must produce a \code{Date},
even when an individual input is itself ambiguous (e.g., contains a time as well as a date).

\paragraph{Recursive agents.}
A careful reader might have noticed that \tool's \code|agent| function does not implement the ``agent loop''---%
the iterative interleaving of LLM reasoning, tool calls, and observation of intermediate results before the next step;
our primitive instead generates a complete program in one shot.
LBAC nonetheless supports the agent loop---and other iterative workflows---%
because agent-generated code can itself call \code{agent},
yielding a \emph{recursive agent}{~\citep{zhang2026recursiveLM}}.
Returning to the literature researcher of \autoref{fig:lbac-overview}:
rather than satisfying the user's query in one shot (as in \autoref{fig:teaser}, right),
the agent might first peek at one search result to check that its keywords are reasonable
and that the DBLP API is operational.
In that case, the top-level \code{agent} call could generate the following program:
\begin{minted}{haskell}
p1 :: BibIO ()
p1 = do hits <- dblpSearch "differential privacy"
        if null hits then return () -- no results
        else do
          bib <- dblpFetchBib (head hits)
          let newPrompt = ... ++ "First search result: " ++ show bib
          agent bibLib newPrompt
\end{minted}
The recursive call carries the observation forward in its prompt,
and the sub-agent decides whether to refine the search or write to the file.

A key consequence of this design is that \emph{policies compose hierarchically}.
The recursive \code{agent} call appears inside a \code{BibIO} \code{do}-block,
so Haskell infers its target type to be \code{BibIO ()},
which means the sub-agent is bound by the same policy as its parent {(\code{()} means that the program does not return a value)}.
A parent may impose \emph{stricter} policies on its children---%
for instance, by giving the recursive call an effect-free target type to forbid further I/O---%
but it cannot relax them: it cannot call \code{agent defs newPrompt :: IO ()},
for the same reason that the top-level agent cannot call \code{writeFile}.

\paragraph{Guarantees{.}}
The core idea of LBAC can be summarized by the following mantra:
\begin{center}
    \emph{If a host language enforces a property for all well-typed programs, and all agent-generated code must typecheck before execution, then that property holds for the entire agentic system.}
\end{center}

For this guarantee to hold, it is essential that not only agent-generated programs, but the entire agentic system---including scaffolding code, control flow, and retry logic---is expressed in the host language.
This is the most important difference from prior approaches:
in LBAC, control is not implemented outside the language, but expressed within it. 
We formalize and prove this claim in \autoref{app:soundness}.

\section{Evaluation}
\label{sec:evaluation}

We evaluate LBAC through three case studies:
data provenance, filesystem sandboxing, and information-flow control.
Each case study follows the same pattern described in \autoref{sec:programming-model}:
(1) define a security property,
(2) encode that property as a restricted effect type, and
(3) expose an API for that type ensuring the property.

\subsection{Data Provenance (continued from \autoref{sec:example})}
\label{subsec:data-provenance}
\textbf{Property:} \code{BibIO a} represents computations whose disk writes only contain data from DBLP and eventually produce a value of type \code{a}.

As discussed in \autoref{sec:example}, LBAC enforces this property by associating authority with types.
Values obtained from DBLP are wrapped in an abstract type
\code{Trusted}, whose constructor is hidden from untrusted code.
As a result, programs cannot forge trusted values from arbitrary
strings; they can only obtain them through approved DBLP APIs.
Effects are similarly restricted.
Output operations are exposed only through the \code{BibIO} interface
and require \code{Trusted} inputs.
Consequently, any successful output produced by a well-typed
\code{BibIO} program must derive from DBLP-approved data.

\subsection{Capability-Based Filesystem Sandboxing}
\label{subsec:filesystem-sandboxing}
\textbf{Property:}
\code{RIO a} represents computations that may only access files within
authorized directories and eventually produce a value of type
\code{a}.

\begin{wrapfigure}{r}{0.57\textwidth}
\vspace{-1.75em}
\begin{minted}{haskell}
example :: Path -> RIO ()
example root = do
  contents <- readRIO (root // "input.txt")
  writeRIO (root // "output.txt") contents
\end{minted}
\vspace{-1.7em}
\end{wrapfigure}
For example, the \code{RIO} program on the right reads a file and writes its
contents to a sibling location.

This behavior is enforced via a \textit{capability} discipline: rather than naming
filesystem locations as raw strings, programs operate on values of type
\code{Path}, which represent unforgeable tokens of authority over a
particular directory subtree.
Possession of a \code{Path} is itself the proof that the holder is
permitted to access that path; without one, no I/O operation can be
expressed in the first place.

In the example, \code{root} is such a token and \code{root // "input.txt"}
narrows it to a specific file.
{Suppose instead we tried to read a file outside \code{root}'s subtree.
Attempting to pass a string containing a filepath directly---e.g.\ \mintinline{haskell}{readRIO "/etc/passwd"}---would
not even typecheck, since \code{readRIO} demands a \code{Path}, not a
\code{String}.}

\begin{wrapfigure}{r}{0.52\textwidth}
\vspace{-1.6em}
\begin{minted}{haskell}
attempt :: Path -> RIO String
attempt root = 
  readRIO (root // "../../etc/passwd")
\end{minted}
\vspace{-1em}
\end{wrapfigure}
\noindent{Moreover, trying to construct a \code{Path} outside \code{root}'s subtree,
as in the example on the right, would fail because \code{//} performs runtime path resolution:}
it follows symbolic links and rejects any result that would fall outside
the original capability's subtree.
Internally, \code{RIO a} is implemented as an opaque type wrapping an \code{IO a} which
is gated behind capabilities. 
The full API is described in \Cref{app:rio-api}.

We argue that enforcing capabilities at the language level provides a clean abstraction for reasoning about filesystem privilege.
Existing agent systems such as Claude Code typically implement sandboxing via two layers: an OS-level sandbox (e.g. \texttt{bubblewrap}), and permission systems which match bash commands against string patterns and/or ask for user approval.
{We next discuss how LBAC compares with each of these approaches.}

\paragraph{LBAC vs. systems-level defenses.}
Our language-based approach is complementary to OS-level defenses, with differing tradeoffs.
Unlike OS sandboxes, which confine a process to a static set of directories, capabilities are first-class values that can be constructed, transferred, and narrowed at runtime.
For example, an agent can accept and return capabilities, or spawn subagents with strictly more restrictive ones.
Capabilities can also generalize beyond physical filesystems to logical resources (for example, allowing access to all files on a git branch).

\paragraph{LBAC vs. permission rules.}
Approaches which are based on processing bash commands as strings have two flaws: they are a poor user-facing abstraction, and treating commands as strings is inherently brittle from a security standpoint.
Conceptually, users typically want to describe a policy about file resources, for example ``Claude can freely modify files in this directory''.
String-based permission systems instead reason at the level of individual commands, which often results in long, opaque approval prompts. 
For example it is not uncommon to see Claude code ask for approval for long bash commands, e.g.:
\begin{minted}{bash}
cd services/platform/identity/auth-provider/packages/api-client && 
sed -i 's/"timeout": 30/"timeout": 60/' config/client.json && python 
scripts/run_integration_tests.py > logs/integration-$(date +%Y%m%d).log 2>&1
\end{minted}

Moreover, string-based enforcement is inherently brittle: it requires faithfully reasoning about every syntactic pattern a bash command can contain.
For example, a regression in Claude Code in March 2026 caused permission bypasses on compound shell invocations joined with \texttt{\&\&}, because only the first clause was matched against the deny list~\cite{ccissue36637}. 
Agents with access to interpreters can also bypass rules entirely with a command like \mintinline{bash}{python -c "open('denied.txt').read()"}, since the matcher has no visibility into the semantics of commands outside the shell. 
LBAC {prevents} the first failure mode because enforcement follows from the type system rather than from heuristics based on parsing syntax.
LBAC {eliminates} the second because the agent already operates within a programming language, and so has no need to call into a separate interpreter to perform computations.

\subsection{Information Flow Control}
\label{subsec:confidentiality}
\textbf{Property:} \code{DC a} represents computations that satisfy information-flow policies and eventually produce a value of type \code{a}.

LBAC integrates with LIO~\cite{DBLP:journals/jfp/StefanMMR17}, a library for coarse-grained information-flow control (IFC). In IFC systems, data carries labels describing how it may propagate through the program. In our setting, we use LIO's disjunction category (DC) labels~\cite{DBLP:conf/nordsec/StefanRMM11} to track two pieces of information: confidentiality (``who may observe this data?''), and
integrity (``what sources influenced this data?'').

\begin{wrapfigure}{r}{0.59\textwidth}
\vspace{-1.6em}
\begin{minted}{haskell}
badWrite :: DC ()
badWrite = do
  content <- httpGet "http://phony.com/data" 
  -- curr. label's integrity now "phony.com"
  writeToUser content -- error
\end{minted}
\vspace{-1em}
\end{wrapfigure}
For example, LIO would throw an error on the example on the right.
Every LIO computation is associated with a floating \emph{current label} that summarizes the labels of all data observed so far. 
Reading from a source raises the current label to incorporate the source's label, so after the \code{httpGet}, the computation's integrity is tainted by \code{phony.com}.
Writing to a sink requires that the integrity of the current label is higher than the integrity of the sink (``user'') so LIO will throw a runtime error.

\paragraph{LBAC generalizes the dual-LLM pattern.}
The dual-LLM pattern of Willison \cite{willison2023duallm} defends against prompt injection by partitioning the agent into a privileged LLM, which has tool access but never sees untrusted data, and a quarantined LLM, which processes untrusted data but has no tool access.
Communication between the two is restricted to opaque variable references, so an injected instruction in untrusted data cannot reach a tool call.

\textit{In LBAC, the dual-LLM pattern is simply a specific use of
existing IFC primitives.} 
The \code{toLabeled} function in LIO runs a sub-computation in isolation, allowing its current label to rise as it observes untrusted data, and returns the result as an opaque \code{Labeled} value without affecting the outer computation's label.
Wrapping a recursive call to \code{agent} in \code{toLabeled} therefore
results in a {\code{Labeled} output, which the privileged LLM cannot directly inspect without raising its label.}
A critical difference from dual-LLM is that \textit{LBAC does not require the subagent to forgo tool use}.
APIs written in LIO naturally gate effectful operations behind checks on the current label of the computation.
Therefore, control-flow integrity follows from the label discipline rather than from a wholesale ban on subagent tool use.

{To illustrate, suppose an agent invokes a subagent inside a \code{toLabeled} block to write a DM.
The subagent can call tools and read untrusted data, but returns the message as a \code{Labeled DCLabel Message}, leaving the outer agent's current label unchanged.
Instead of inspecting the message, the outer agent passes the labeled result to \code{sendDM}, a tool that operates directly on labeled data and enforces policies on the recipient and message contents:}
\begin{minted}{haskell}
sendDM :: User -> Labeled DCLabel Message -> DC ()
\end{minted}

{Like the privileged agent in the dual-LLM pattern, the outer agent in our example can send the message without being exposed to untrusted data.
However, unlike the dual-LLM pattern, which forbids all tool calls by a quarantined LLM, \tool allows the subagent to call tools because information flow control restricts tool calls after the subagent reads untrusted data.}

\paragraph{Comparison to CaMeL}
\label{sec:eval:camel}
We compare \tool to CaMeL on the Slack suite of
AgentDojo~\cite{agentdojo}, which measures \emph{utility} (fraction of benign
tasks completed) and \emph{security} (fraction of injection attempts
resisted).  
The Slack suite consists of 21 user tasks and 5 injection tasks.
Attacks in AgentDojo use every possible (user task, injection task) pair for a total of 105 possible tasks.
For baseline comparisons that do not involve attacks, we only use user tasks.
{
{}
We ran our experiments with Opus 4.7, Opus 4.8, GPT 5.4, and Sonnet 5.
}
For both tools, we report pass@1.

As \tool is implemented in Haskell and AgentDojo is written in Python, we
connect the two through inter-process communication. 
Both \tool and CaMeL can be run with or without IFC policies enabled.
Without policies, CaMeL still uses its dual-LLM architecture but performs no {dataflow} checks.
\tool likewise has two configurations: an unsafe API where the
Slack API {uses} plain \code{IO}, and a policy-enforcing version where the API {uses} \code{LIO}.
{
{}
LIO primitives enable us to define a novel style of policy based on the following principle: whenever a model reads untrusted data, its privilege should lower to the privilege of the data's source. This is an example of a ``low watermark''~\citep{biba1977integrity} policy. An interesting direction for future work that \tool enables is to standardize benchmark policies across agent implementations using existing IFC systems such as LIO, rather than tying policies to each agent's bespoke IFC system.
}

\paragraph{Findings.}
{
{}
\tool achieves higher utility than CaMeL across nearly all models and attack settings, completing roughly twice as many tasks with policies enforced and $\sim$15\% more without (Table~\ref{tab:agentdojo-utility} {in \hyperref[app:agentdojo-evaluation]{Appendix~\ref*{app:agentdojo-evaluation}}}). \tool's utility failures without policies are mostly due to model refusals.

For both CaMeL and \tool, all evaluated models resisted all 105 AgentDojo attacks even without policy enforcement. However, prompt injection attacks are becoming increasingly sophisticated: for example, GPT-Red, OpenAI's automated red-teaming system, consistently circumvented GPT-5.5's defenses~\citep{openai2026gptred}. \citet{nasr2026attacker} and \citet{zhan2025adaptive} argue for evaluating defenses against \emph{adaptive} attacks: motivated by this, we simulate \emph{worst-case} injections by forcing compromised subagents to execute attacker instructions.

\paragraph{Adversarial evaluation.}
In \tool, all data is read through recursive calls to \texttt{agent}, which receive prompts of the form \texttt{Task: \{task\} Input: \{input data\}}. We modified \tool's code so that whenever a subagent receives input data containing the string \texttt{TODO:} (the AgentDojo direct injection format), the prompt is replaced with only the input data. This simulates a \emph{worst-case} injection: the subagent executes the attacker's instructions directly, rather than merely being exposed to them. Table~\ref{tab:adversarial-security} reports our results.

\setcounter{table}{1}
\begin{table}[htbp]
{
  \centering
  \small
  \caption{Adversarial evaluation: injections resisted and tasks completed under attack (higher is better).}
  \label{tab:adversarial-security}
  \setlength{\tabcolsep}{6pt}
  \begin{tabular}{llrr}
    \toprule
    {System} & {Model} & {Injections resisted}{\footnotemark} & {Tasks completed} \\
     & & {(/735)} & {(/735)} \\
    \midrule
    \tool & Opus 4.7 & 735 & 351 \\
     & Opus 4.8 & 735 & 312 \\
     & GPT 5.4 & 735 & {338} \\
     & Sonnet 5 & 735 & 294 \\
    \midrule
    \tool (no policy) & Opus 4.7 & 425 & 523 \\
     & Opus 4.8 & 460 & 557 \\
     & GPT 5.4 & {621} & {558} \\
     & Sonnet 5 & 540 & 494 \\
    \bottomrule
  \end{tabular}
  \par
}
\end{table}
\footnotetext{\label{fn:agentdojo-scoring}{AgentDojo has a scoring false positive in one injection task which affects {29} out of the 2940 runs: we manually reviewed all {29} cases and found no security violations. See \hyperref[app:adversarial-evaluation]{Appendix~\ref*{app:adversarial-evaluation}} for details.}}

Without policies, up to 42\% of injections succeed depending on the model. With policies enforced, \tool resists all injections.

{Enforcing the policies reduces task utility because the policies may also rule out legitimate agent behavior (Table~\ref{tab:adversarial-security}). Even with high false positive rates, sound deterministic defenses remain useful as conservative filters because they have \emph{zero} false negatives relative to the enforced policy. As \citet{kolluri2026optimizingagentplanningsecurity} argue, in human-in-the-loop systems deterministic defenses can be used to safely reduce the need for user approvals.}

Systems implementing the dual LLM pattern such as CaMeL would also prevent prompt injection by making tool calls independent of data. However, this separation reduces utility as shown in Table~\ref{tab:agentdojo-utility} {in \hyperref[app:agentdojo-evaluation]{Appendix~\ref*{app:agentdojo-evaluation}}}.
}

\section{Related Work}
\label{sec:related-work}

\paragraph{{Language-based approaches to prompt injection.}}

\looseness=-1
The closest work to LBAC is on language-based approaches to preventing prompt injection.
{
{Fides} and {Invariant}~\citep{costa2025securingaiagentsinformationflow,balunovic2024formal} use IFC to enforce confidentiality and integrity policies on tool calls.
{Prudentia}~\citep{kolluri2026optimizingagentplanningsecurity} builds on Fides by allowing interactive endorsement of data and approval of individual actions.
{GAAP}~\citep{stanley2026aiagentexecutionenvironment} is an agent execution environment that uses IFC to protect private data even from the model provider.
{TACIT~\citep{odersky2026trackedcapabilities} uses Scala's capture checking to enforce capabilities and a limited form of IFC on agent-generated code.}
{LLMbda~\citep{garby2026llmbda} adds agent primitives to a basic information-flow calculus, with a formal proof of noninterference.}

\looseness=-1
{An essential design principle of LBAC that separates it from these prior approaches is to decouple} the agent implementation from the policy enforcement mechanism.
{This approach has two benefits: (1) LBAC enables whole-system policies that cover both agents (as in prior approaches) \emph{and} the surrounding application code; and (2) it allows direct reuse of} established systems such as LIO~\citep{DBLP:journals/jfp/StefanMMR17}.
LIO has many features (global process clearance, privileges, concurrency, {labels which track both integrity and confidentiality}), {many of which these systems re-implement parts of.
Using LIO removes the need to trust a custom IFC implementation as there is a long line of work developing LIO and formal soundness proofs of correctness~\citep{DBLP:conf/haskell/StefanRMM11,DBLP:journals/jfp/StefanMMR17}.}
}

{
{}
{Another approach taken by work such as CaMeL~\citep{debenedetti2025defeatingpromptinjectionsdesign} and Guardians of the Agents~\citep{DBLP:journals/cacm/Meijer26,amin2026guardiansrepo} is to generate a program upfront and directly apply language-based techniques (IFC and SMT verification) to the program.}
}
An important difference, however, is that both lines of work additionally impose an \textit{architectural constraint} of \emph{code and data separation}:
the agent is supposed to generate code upfront, before reading any data,
which severs the information flow from potentially malicious prompts to the code's control flow.
By contrast, our approach allows the agent to interleave reading data
with calling tools, with security enforced by the type system rather than by
architectural separation.

{
{Other applications of language-based techniques include} {PASS}~\citep{crouse2024formally} and {Safety Chip}~\citep{yang2023safetychip}{, which} use Linear Temporal Logic (LTL) formulas to restrict the allowed execution traces of agents, via constrained decoding and runtime checks respectively.
}

{}

{
\paragraph{Systems-level defenses.}
Network and filesystem sandboxes~\citep{bubblewrap,merkel2014docker} are commonly used by production systems and can complement language-based defenses.
In general, compared to language-based approaches, which can attenuate policies based on application logic, systems-level defenses are more coarse-grained.
{MiniScope}~\citep{zhu2025miniscopeprivilegeframeworkauthorizing} is a framework for restricting tool-calling agents to user-authorized permissions.
{ceLLMate}~\citep{meng2026cellmatesandboxingbrowserai} is a browser sandbox that enforces policies at the HTTP request level.
Chrome's {Agent Origin Sets}~\citep{parker2025agentorigins} restrict which web origins agents can read or modify.
}

{LLM-based defenses such as Conseca~\citep{tsai2025contextualagentsecuritypolicy} and IsolateGPT~\citep{wu2025isolategpt} rely on separate language models to generate policies which describe permissible actions and data accesses.}

\paragraph{{Programming-language foundations.}}
\label{sec:rw-pl}

\tool builds on existing programming-language mechanisms rather than introducing new type machinery.
The \code|RIO|, \code|TIO|, and \code|LIO| monads of \autoref{sec:programming-model}
are direct instances of \emph{language-based security}---%
the use of type systems and runtimes to enforce information-flow and capability policies---%
following the lineage of
\citet{DBLP:journals/jsac/SabelfeldM03},
\citet{DBLP:conf/dagstuhl/SchneiderMH01},
\textsc{Jif}~\citep{DBLP:conf/popl/Myers99},
Safe Haskell~\citep{DBLP:conf/haskell/TereiMJM12},
LIO~\citep{DBLP:conf/haskell/StefanRMM11,DBLP:journals/jfp/StefanMMR17}, and work on effect systems~\citep{DBLP:conf/esop/PlotkinP09,DBLP:conf/popl/Leijen17}.

\section{Conclusion}

We presented LBAC, a programming model for LLM agents that treats agent control as a type- and runtime-enforcement problem inside a unified host language.
LBAC allows agents to interleave reasoning, tool use, and code generation dynamically,
while enforcing information-flow and capability policies over the resulting execution.
Our central observation is that mechanisms proposed for secure and controllable agents
can be expressed naturally using existing programming-language techniques once generated code and scaffold execute inside the same typed runtime.
This removes the need for architectural workarounds such as strict code/data separation.

More broadly, LBAC suggests that agent control is fundamentally a programming-languages problem.
Richer type disciplines---including refinement types~\citep{DBLP:conf/icfp/VazouSJVJ14,DBLP:journals/pacmpl/PolikarpovaSYIH20,DBLP:journals/pacmpl/LehmannGVJ23},
linear types~\citep{DBLP:journals/pacmpl/BernardyBNJS18},
amortized resource analysis~\citep{DBLP:conf/popl/HoffmannAH11},
and graded types~\citep{DBLP:conf/pldi/KnothWP019}---%
could naturally extend our LBAC implementation with guarantees such as bounded resources,
one-shot authorizations, and termination properties.
\begin{ack}
{The project is supported in part by a Microsoft Faculty Fellowship and NSF under grants CCF-2422214, CCF-2506134, CCF-2446711. Any opinions, findings, and conclusions or recommendations expressed in this publication are those of the authors, and do not necessarily reflect the views of the sponsoring entities.}

{Loris D'Antoni holds concurrent appointments as a Professor at the University of California San Diego and as an Amazon Scholar. This paper describes work performed at the University of California San Diego and is not associated with Amazon.}
\end{ack}

\bibliographystyle{plainnat}
\bibliography{reference}

\newpage
\appendix
\crefalias{section}{appendix}

\section{{{}Evaluation Data}}
\label{app:data}

{The code for LBAC, our experiments, our policies, and the raw data are available in our public GitHub repository: {\url{https://github.com/timothytmzhou/elba/tree/neurips}}.}

{
{}
}

{
\subsection{AgentDojo utility and security}
\label{app:agentdojo-evaluation}
}
\setcounter{table}{0}
\begin{table}[htbp]
{
  \centering
  \small
  \caption{Utility on the AgentDojo Slack suite (tasks completed; higher is better).}
  \label{tab:agentdojo-utility}
  \setlength{\tabcolsep}{4pt}
  \begin{tabular}{llrrrr}
    \toprule
    {System} & {Model} & {No Attack} & {Direct} & {Important Instr.} & {Tool Instr.} \\
     & & {(/21)} & {(/105)} & {(/105)} & {(/105)} \\
    \midrule
    \tool & Opus 4.7 & 10 & 52 & 51 & 50 \\
     & Opus 4.8 & 9 & 49 & 47 & 45 \\
     & GPT 5.4 & 10 & 51 & 48 & 49 \\
     & Sonnet 5 & 8 & 43 & 46 & 40 \\
    \midrule
    \tool (no policy) & Opus 4.7 & 18 & 88 & 84 & 92 \\
     & Opus 4.8 & 16 & 93 & 90 & 92 \\
     & GPT 5.4 & 16 & 84 & 80 & 79 \\
     & Sonnet 5 & 14 & 78 & 78 & 79 \\
    \midrule
    CaMeL & Opus 4.7 & 5 & 24 & 23 & 22 \\
     & Opus 4.8 & 6 & 30 & 29 & 30 \\
     & GPT 5.4 & 5 & 23 & 26 & 25 \\
     & Sonnet 5 & 4 & 25 & 23 & 23 \\
    \midrule
    CaMeL (no policy) & Opus 4.7 & 15 & 69 & 69 & 73 \\
     & Opus 4.8 & 16 & 78 & 82 & 82 \\
     & GPT 5.4 & 17 & 82 & 82 & 84 \\
     & Sonnet 5 & 13 & 56 & 58 & 55 \\
    \bottomrule
  \end{tabular}
  \par
}
\end{table}
\setcounter{table}{2}

{
We ran CaMeL in both ``strict'' and ``normal'' mode, and found essentially no difference (all cells within 4\% of each other). We report strict mode.
}

\subsection{{Confidence intervals}}
\label{app:confidence-intervals}
{Table~\ref{tab:agentdojo-cis} reports 95\% confidence intervals for the difference in task completion rates between \tool and CaMeL. We compare the two systems on the same tasks and, under attack, the same injections. Table~\ref{tab:adversarial-utility-cis} reports 95\% confidence intervals for task completion rates {and injection resistance rates without policies} in the adversarial evaluation.} {We compute bootstrap confidence intervals using the results from the seven benchmark runs.}

\begin{table}[H]
\centering
\small
\caption{95\% confidence intervals for the difference in task completion rates ($\tool - \text{CaMeL}$), in percentage points, computed using Newcombe's score interval for the difference in paired proportions.}
\label{tab:agentdojo-cis}
\setlength{\tabcolsep}{3pt}
\begin{tabular}{llrrrr}
\toprule
Model & Policies & No attack & Direct & Important Instr. & Tool Instr. \\
\midrule
Opus 4.7 & Yes & $[+2.9, +41.8]$ & $[+17.3, +35.3]$ & $[+17.4, +35.3]$ & $[+17.4, +35.3]$ \\
Opus 4.8 & Yes & $[-7.4, +34.1]$ & $[+9.9, +25.8]$ & $[+8.7, +25.2]$ & $[+6.3, +22.0]$ \\
GPT 5.4 & Yes & $[+2.9, +41.8]$ & $[+17.4, +35.3]$ & $[+12.0, +29.4]$ & $[+13.6, +31.5]$ \\
Sonnet 5 & Yes & $[-4.7, +40.4]$ & $[+7.9, +26.0]$ & $[+12.4, +30.9]$ & $[+7.9, +24.3]$ \\
\midrule
Opus 4.7 & No & $[-8.4, +35.9]$ & $[+8.0, +27.9]$ & $[+4.3, +24.0]$ & $[+9.6, +26.7]$ \\
Opus 4.8 & No & $[-16.9, +16.9]$ & $[+6.4, +22.5]$ & $[+0.5, +15.1]$ & $[+2.5, +17.0]$ \\
GPT 5.4 & No & $[-24.0, +14.5]$ & $[-5.5, +9.3]$ & $[-8.8, +4.9]$ & $[-11.4, +1.7]$ \\
Sonnet 5 & No & $[-22.7, +31.3]$ & $[+11.2, +30.1]$ & $[+8.4, +29.0]$ & $[+12.8, +32.2]$ \\
\bottomrule
\end{tabular}
\end{table}

\begin{table}[H]
\centering
\small
\caption{Task completion rates and 95\% confidence intervals for the adversarial evaluation.}
\label{tab:adversarial-utility-cis}
\begin{tabular}{llrcrc}
\toprule
\multirow{2}{*}{Model} & \multirow{2}{*}{Policies} & \multicolumn{2}{c}{{Tasks completed (\%)}} & \multicolumn{2}{c}{{Injections resisted (\%)}} \\
\cmidrule(lr){3-4}\cmidrule(lr){5-6}
 & & {Rate} & {95\% CI} & {Rate} & {95\% CI} \\
\midrule
Opus 4.7 & Yes & 47.8 & [46.1, 49.3] & {100.0} & {N/A} \\
Opus 4.8 & Yes & 42.4 & [41.4, 43.4] & {100.0} & {N/A} \\
GPT 5.4 & Yes & 46.0 & [44.2, 47.6] & {100.0} & {N/A} \\
Sonnet 5 & Yes & 40.0 & [38.6, 41.4] & {100.0} & {N/A} \\
\midrule
Opus 4.7 & No & 71.2 & [70.1, 72.2] & {57.8} & {[56.3, 59.3]} \\
Opus 4.8 & No & 75.8 & [74.6, 77.0] & {62.6} & {[62.2, 63.1]} \\
GPT 5.4 & No & 75.9 & [73.9, 78.0] & {84.5} & {[82.4, 86.9]} \\
Sonnet 5 & No & 67.2 & [65.0, 69.3] & {73.5} & {[72.8, 74.1]} \\
\bottomrule
\end{tabular}
\end{table}

{For the Table~\ref{tab:agentdojo-utility} experiments, both systems resisted all 105 injections in every model and attack setting, including without policies. For Table~\ref{tab:adversarial-security}, security with policies is enforced deterministically, so it is not a statistical quantity.}

{
\subsection{Adversarial evaluation scoring}
\label{app:adversarial-evaluation}
AgentDojo has a scoring false positive in one injection task which affects {29} out of 2940 runs: it counts accessing a static URL as a successful injection. Our policies do not prevent accessing the URL itself as it is not a security violation. Our policies would deny a request whose URL contained a parameter leaking data, a request made after the agent had accessed sensitive data (because requests may exfiltrate information about the LLM's context), or a subsequent privileged action based on the page's contents (due to the possibility of prompt injections). We manually reviewed all {29} cases and found no security violations.
}

{
\subsection{Retries and runtime}
\label{app:agentdojo-retries}
\tool retries code generation when an agent produces a program that fails to typecheck. Retries apply only to typechecking errors, which occur before any tool calls execute, and not runtime errors. Agent calls required an average of 0.1 retries and a maximum of 3 retries. A run may contain multiple agent calls. If the LLM repeatedly fails to produce well-typed code, a runtime error is thrown. This happened in 0.26\% of runs. LIO may also throw runtime errors because it is a dynamic information-flow control system (policies are not checked statically).
}

\begin{table}[htbp]
{
  \centering
  \small
  \caption{Number of retries on AgentDojo.}
  \label{tab:agentdojo-retries}
  \setlength{\tabcolsep}{4pt}
  \begin{tabular}{lrrr}
    \toprule
    {Model} & {Avg retries} & {Runs exceeding} & {\% of time} \\
     & {per run} & {retry limit} & {on retries} \\
    \midrule
    opus-4.7 & 0.05 & 0/672 & 0.7\% \\
    opus-4.8 & 0.14 & 5/672 & 5.0\% \\
    sonnet-5 & 0.14 & 2/672 & 2.3\% \\
    gpt-5.4 & 0.19 & 0/672 & 4.4\% \\
    \midrule
    all & 0.13 & 7/2688 & 3.4\% \\
    \bottomrule
  \end{tabular}
  \par
}
\end{table}

{
The total runtime of \tool on AgentDojo is about $3\times$ that of CaMeL's: this is primarily due to the fact that CaMeL generates a single program upfront, whereas \tool agents perform additional LLM calls to generate programs in response to data (which increases expressiveness, and ultimately, utility).
}

{
\subsection{BibTeX evaluation}
\label{app:bibtex-evaluation}
We evaluated the BibTeX data provenance case study from \autoref{subsec:data-provenance}. The evaluation consisted of 100 tasks where an agent must search the web and write a BibTeX file of citations for a given computer science topic. A citation is hallucinated if its authors, year, or DOI are incorrect. We compared (1) an agent implemented in \tool using opaque types to enforce data provenance and (2) a vanilla \texttt{pydantic\_ai} agent with a web search tool. Both agents used Sonnet 5. The pydantic-ai agent writes the citations itself from search results and memory, so it can invent details such as DOIs. As expected (Table~\ref{tab:bibtex-evaluation}), \tool prevents citation hallucinations.
}
\begin{table}[htbp]
{
  \centering
  \small
  \caption{BibTeX evaluation (hallucinated citations; lower is better).}
  \label{tab:bibtex-evaluation}
  \setlength{\tabcolsep}{4pt}
  \begin{tabular}{lr}
    \toprule
    System & Hallucinated Citations \\
    \midrule
    pydantic-ai agent + web search tool & 19/1002 \\
    \tool & 0/925 \\
    \bottomrule
  \end{tabular}
  \par
}
\end{table}

{
The tool-calling baseline also used about $4\times$ more LLM calls, $2.5\times$ more output tokens, and $65\times$ more input tokens.
}

\section{Soundness of LBAC}
\label{app:soundness}

In this appendix we make precise the central soundness claim of LBAC---%
\emph{any property the host type system enforces for well-typed programs is also enforced for programs containing} \code{agent} \emph{calls, provided each agent-generated program typechecks before execution}---%
and prove it by reduction to the type soundness of the host language.

\subsection{{Host Language Properties}}
\label{app:host}

We work parametrically in a host programming language $L$ {supporting standard features like conditionals, functions, and recursion}.
$L$ has {expressions denoted $e$} and types $t$,
and a typing judgment $\Gamma \vdash_L e : t$ where $\Gamma$ is a typing environment {assigning types to variables}.

$L$ has an operational semantics given by a small-step reduction relation $\sigma \to_L \sigma'$ over \emph{configurations}.
A configuration {$\sigma=\langle e,\kappa\rangle$} combines a remaining program {$e$} with some auxiliary state {$\kappa$} (e.g., a store, a heap, an I/O history, a security label).
{An \emph{execution} is a possibly infinite sequence of configurations of the form
\[
a_1 \to_L a_2 \to_L a_3 \to_L \cdots.
\]
}
{A \emph{trace} is the sequence of tool calls an execution makes which produce an observable side effect.
The observable behavior of a program $e$ is completely characterized by the set of traces it induces ($\operatorname{traces}(e)$).}

{A \emph{type system property} for a type $t$ is a predicate $P$ on sets of traces such that
\[
\forall e \in L.\quad \vdash_L e : t \;\Longrightarrow\; P(\operatorname{traces}(e)).
\]
}

\subsection{{Semantics of Agents}}
\label{app:extension}

{Given a language $L$, we define a new language $L_{\mathsf{agent}}$ which extends $L$ with a new primitive $\text{\code{agent}}_t^n$.
Here, every agent call is annotated with two pieces of data: a return type $t$ and a bound $n$ on the depth of recursive subagent calls it can make.
Languages such as Haskell that implement type inference can determine $t$ from context and the depth is usually a parameter supplied by, e.g., a configuration file rather than written explicitly by the programmer.}

{The typing rules for $L_{\mathsf{agent}}$ are the same as $L$, except $\text{\code{agent}}_t^n$ is given the type:
\[
\text{\code{agent}}_t^n : \mathsf{Defs} \to \mathsf{Prompt} \to \mathsf{Option}\;t
\]
where $\mathsf{Defs}$ represents a set of definitions the agent can use in the code it constructs, $\mathsf{Prompt}$ is a string containing the agent's prompt, and the output type is $\mathsf{Option}\;t$ as the agent may fail to produce a result.}

{We model LLMs in a ``black box'' manner: we assume there is some arbitrary computable {oracle} function $\mathcal{O}_t$ which, given a depth $n$, a set of definitions $\mathsf{defs}$, and a prompt $\mathsf{prompt}$, outputs a program whose agent calls have depth at most $n$.}
{While we assume that the function is completely arbitrary, we do not model the nondeterminism of the language model.
Nondeterminism is irrelevant to the properties we consider in this paper, such as those in \autoref{sec:evaluation}.
For cases where determinism is required, one can implement a wrapper that caches the results of prompts to a language model.}

{The semantics of $L_{\mathsf{agent}}$ is the same as the semantics of $L$, but with an additional rule to execute agent calls.
An $\text{\code{agent}}_t^n$ expression evaluates by querying the oracle for a program which may contain agent calls up to depth $n-1$.
If the program typechecks under the given definitions, the agent expression evaluates to that program wrapped in $\mathsf{Some}$ to indicate success.
Otherwise it evaluates to $\mathsf{Nothing}$.}
{$\Gamma_{\mathsf{defs}}$ is a typing environment for the given definitions.}
{
\[
\frac{\mathcal{O}_t(n,\mathsf{defs},\mathsf{prompt})=e_{\mathsf{gen}} \quad \Gamma_{\mathsf{defs}}\vdash e_{\mathsf{gen}}:t}
{\langle\text{\code{agent}}_t^{n+1}\;\mathsf{defs}\;\mathsf{prompt},\kappa\rangle
\;\to\;
\langle\mathsf{Some}(e_{\mathsf{gen}}),\kappa\rangle}
\]}
{
\[
\frac{}
{\langle\text{\code{agent}}_t^n\;\mathsf{defs}\;\mathsf{prompt},\kappa\rangle
\;\to\;
\langle\mathsf{Nothing},\kappa\rangle}
\quad\text{otherwise}
\]}

{We write $o_t$ for an implementation of $\mathcal{O}_t$ in $L$.
$o_t$ calls an LLM and parses its output into a datatype representing syntax trees.
We assume $L$ has a function $\mathsf{eval}_t$ which, given an element of the syntax tree datatype and a target type $t$, typechecks the tree and if successful evaluates it into a value of type $\mathsf{Option}\;t$.}

\label{app:theorem}
{
\begin{theorem}[LBAC Soundness]
\label{thm:lbac-soundness}
Every type system property for $L$ is a type system property for $L_{\mathsf{agent}}$.
\end{theorem}
}

\begin{proof}
{Let $P$ be a type system property for $L$ at type $t$.
Let $e$ be a program in $L_{\mathsf{agent}}$ with $\vdash_{L_{\mathsf{agent}}} e:t$.
Our goal is to show that $P(\operatorname{traces}(e))$ holds as well.

The main idea is to show the existence of a program $e_L$ in $L$ satisfying:
\begin{enumerate}
    \item $\operatorname{traces}(e_L)=\operatorname{traces}(e)$
    \item $\vdash_L e_L:t$
\end{enumerate}
Then since $P$ is a type system property for $L$, we must have $P(\operatorname{traces}(e_L))$ which implies $P(\operatorname{traces}(e))$.}

{At a high level, we will replace every $\text{\code{agent}}_t^n$ call in the $L_{\mathsf{agent}}$ program $e$ with a call to an $L$ term $a_t^n$ which uses $o_t$ to implement the oracle call.
The main difficulty is that the oracle may return programs containing further agent calls which themselves need to be substituted.
We therefore define $a_t^n$ recursively as follows:
\[
\begin{aligned}
a_t^0\;\mathsf{defs}\;\mathsf{prompt}
&=\mathsf{Nothing},\\
a_t^{n+1}\;\mathsf{defs}\;\mathsf{prompt}
&=\begin{cases}
{\mathsf{eval}_t}\bigl(e_{\mathsf{gen}}[\text{\code{agent}}_s^m\mapsto a_s^m]\bigr)
&\text{if }\Gamma_{\mathsf{defs}}\vdash e_{\mathsf{gen}}:t,\\
\mathsf{Nothing}&\text{otherwise},
\end{cases}\\
&\text{where }e_{\mathsf{gen}}=o_t(n,\mathsf{defs},\mathsf{prompt}).
\end{aligned}
\]
The substitution replaces every $\text{\code{agent}}_s^m$ expression in $e_{\mathsf{gen}}$ with $a_s^m$.
The term $o_t$ by assumption ensures $m\leq n$, so the definition of $a_t^{n+1}$ uses only terms already defined at smaller depths.}

{Next, we define $e_L$ as:
\[
e_L=e[\text{\code{agent}}_s^m\mapsto a_s^m\mid\text{\code{agent}}_s^m\text{ occurs in }e].
\]
Every $a_s^m$ is an $L$ term with the same type as $\text{\code{agent}}_s^m$, so $e_L$ is an $L$ program with $\vdash_L e_L:t$.}

{Finally, we prove that $\operatorname{traces}(e)=\operatorname{traces}(e_L)$.
It suffices to show that replacing any $\text{\code{agent}}_t^n$ call with a call to $a_t^n$ preserves the traces of the program containing the call.
We proceed by induction on $n$ for all result types $t$.
At depth zero, $a_t^0$ and $\text{\code{agent}}_t^0$ return $\mathsf{Nothing}$.
For depth $n+1$, assume the claim holds at all depths at most $n$.
Since $o_t$ computes $\mathcal{O}_t$, we obtain the same generated program $e_{\mathsf{gen}}$ in the calls to $a_t^{n+1}$ and $\text{\code{agent}}_t^{n+1}$.
Each call returns $\mathsf{Nothing}$ if $e_{\mathsf{gen}}$ does not typecheck.
Otherwise, every generated call has depth at most $n$, so the induction hypothesis applies to every substitution in the recurrence.}
\end{proof}

\section{\code{RIO} API}
\label{app:rio-api}
\begin{itemize}
    \item \code{evalRIO :: String -> RIO a -> IO a} is the entry point to the API: given a root directory and an \code{RIO} computation, it runs the computation bound to a capability for that root.
    \item \code{Path} is an opaque type whose constructors are not exported: the only way to obtain a \code{Path} is via the API, ensuring that every \code{Path} in scope was derived from one originally granted by \code{evalRIO}.
    \item \code{// :: Path -> String -> Path} is a binary operator which narrows an existing capability: given a \code{Path} for \code{p} and a string \code{s}, it returns a \code{Path} for \code{p/s}. Crucially, \code{//} performs runtime path resolution: it follows symbolic links and rejects any result that would fall outside the original capability's subtree.
    \item \code{readRIO :: Path -> RIO String}, \code{writeRIO :: Path -> String -> RIO ()}, and \code{ls :: Path -> RIO [Path]} perform reads, writes, and directory listings, each requiring a \code{Path} as proof of authority.
\end{itemize}

\section{System Prompt}
\label{app:prompt-system}
\inputminted[breaklines, fontsize=\small]{md}{prompts/system.md}

\end{document}